\input{aipcheck}

\documentclass[
    ,final            
  ]
  {aipproc}

\layoutstyle{6x9}

\usepackage{natbib}

\begin{document}

\title{Uncovering Local Absorbed Active Galactic Nuclei with Swift and Suzaku}


\author{Lisa M. Winter}{
  address={Hubble Fellow, University of Colorado at Boulder}
}


\begin{abstract}
Detection of absorbed active galactic nuclei and their properties remains an elusive and important problem in understanding the evolution and activation of black holes.  With the very hard X-ray survey conducted by Swift's Burst Alert Telescope -- the first all-sky survey in 30 years -- we are beginning to uncover the characteristics of obscured AGN.  The synergy between Suzaku and Swift has been crucial in pinning down the X-ray properties of newly detected heavily obscured but bright hard X-ray sources.  We review the X-ray and optical spectroscopic properties of obscured AGN in the local Universe, as detected in the Swift survey.  We discuss the relative distribution of absorbed/unabsorbed sources, including ``hidden'' and Compton thick AGN populations.  Among the results from the survey, we find that absorbed AGN are less luminous than unabsorbed sources.  Optical spectra reveal that sources with emission line ratios indicative of LINERs/H II galaxies/composites are the least luminous objects in the sample, while optical absorbed and unabsorbed Seyferts have the same luminosity distributions.  Thus, the least luminous sources are likely accreting in a different mode than the Seyferts.
\end{abstract}

\maketitle


\section{Introduction}

After over sixty years of research on active galaxies, there are still many unanswered questions remaining about the nature of black holes and their connection to the host galaxies.  We do not yet understand why some black holes are active, while others are not, or whether all super massive black holes undergo a phase of activity at some point in their evolution.  If all black holes do undergo an active phase, how does this affect the formation and evolution of the host galaxy?  Further, at an even more basic level, what are the properties of active galaxies?

Selection methods of active galaxies present a major challenge in determining both the properties of active galactic nuclei (AGN) and their host galaxies
\citep{2004ASSL..308...53M}.  Optical and soft X-ray surveys are greatly affected by dust and gas in the line of sight and therefore miss heavily obscured AGN.  In the infrared, the AGN signature is difficult to disentangle from star-formation.  The very hard X-rays, at energies above 10\,keV, are an unambiguous indicator of AGN emission that is unaffected by all but the highest, Compton-thick (N$_{\rm H} > 1.4 \times 10^{24}$\,cm$^{-2}$), levels of obscuration (see Figure 1).  Therefore, the very hard X-rays offer an excellent selection method for determining the basic properties of AGN.

With the launch of the Swift Gamma-ray burst satellite in 2004, the largest all-sky survey in the very hard X-rays is being conducted with the Burst Alert Telescope (BAT; sensitive in the 14--195\,keV band).  The current survey includes over 1000 sources, with more than 600 AGN (Tueller, in prep).  However, the most well-studied sources from the Swift survey are the 102, $|b| > 15^{\circ}$, AGN detected in the 9-month catalog 
\citep{2007arXiv0711.4130T}.  These sources are bright and nearby ($\langle z \rangle = 0.03$), making them excellent targets for multi-wavelength observational studies.  The X-ray \citep{2009ApJ...690.1322W} and optical \citep{2010ApJ...710..503W} spectroscopic properties of these sources are the subject of this article.  In particular, X-ray follow-ups with the Suzaku satellite, which obtains simultaneous coverage in the soft through very hard X-rays, have been crucial in determining the properties of the most heavily obscured AGN detected with Swift.

\begin{figure}
  \includegraphics[height=.35\textheight]{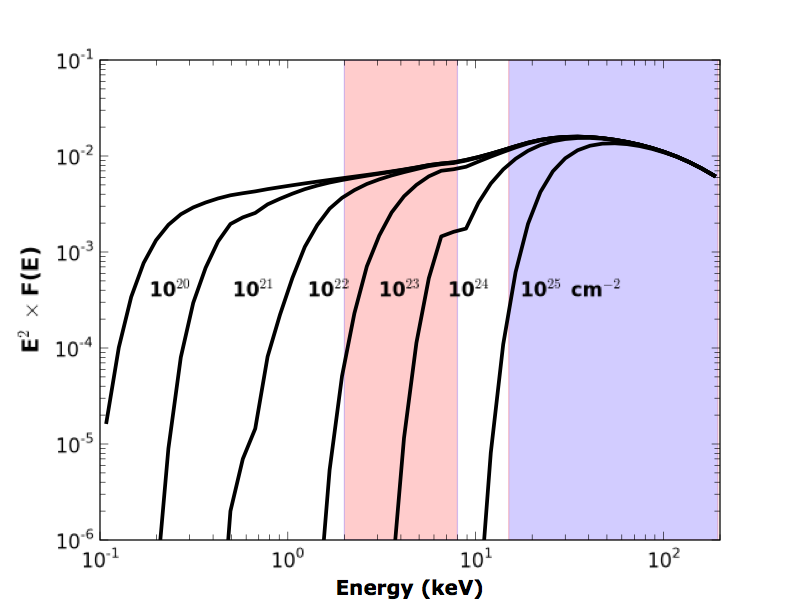}
  \caption{The direct power-law + reflection spectrum of a typical AGN is shown, with different levels of obscuration.  In the Chandra hard X-ray band (2--8\,keV; in red), surveys will miss AGN with moderate levels of obscuration.  For instance, an AGN obscured by a column of $10^{23}$\,cm$^{-2}$ is difficult detect in this band.  The Swift BAT band (14--195\,keV; in blue) is less biased towards obscuration and can easily select Compton-thin (N$_{\rm H} < 10^{24}$\,cm$^{-2}$) AGN. This makes the Swift-detected AGN an excellent sample for determining the basic properties of AGN.}
\end{figure}

\section{The X-ray Spectral Properties of Swift-detected Active Galaxies}

A combination of archival and follow-up X-ray spectroscopic observations in the 0.3--10\,keV band were analyzed for all of the Swift BAT-detected AGN \citep{2009ApJ...690.1322W}.  We fit available spectra from Suzaku, ASCA, XMM-Newton, Chandra, and the Swift XRT to determine the basic properties of our sample.  We found that 45\% of our sources are well-fit by a simple absorbed power-law model (similar to the spectra shown in Figure 1).  The additional 55\% of our sources required a more complex model, which we fit with a partial covering absorber model.  In the partial covering absorber model, the absorber is assumed to cover only a fraction of the direct emission.  We found that all of the sources with the highest measured column densities require the more complex model.  The least absorbed sources, meanwhile, are well fit by the simple model.

An intriguing result from our X-ray analysis was that the fraction of absorbed sources changes with luminosity.  In Figure 2, we show that there are fewer absorbed (N$_{\rm H} > 10^{22}$\,cm$^{-2}$) sources at the highest luminosities (30\%).  However, more than 80\% of the low luminosity AGN are obscured.  A similar result was also found at higher redshift in the Chandra-selected AGN samples 
\citep{2003ApJ...598..886U,2003ApJ...596L..23S,2005AJ....129..578B}.  We found that this relationship also exists in accretion rate, where there are more absorbed sources at the lowest accretion rates.  This result provides a challenge to the unified model.  In the unified model 
\citep{1993ARA&A..31..473A}, both obscured and unobscured AGN are fundamentally the same.  The difference in the observed spectra is caused simply by our viewing angle to the torus.  However, if absorbed sources are less luminous/accreting at a lower rate, as recent results suggest, there must be a modification to this simple model to account for differences in accretion rates.  In the following section, we discuss further clues to this problem from optical spectroscopic follow-ups. 

\begin{figure}
  \includegraphics[trim = 1in 0.8in 1.5in 2.0in, clip, height=.4\textheight]{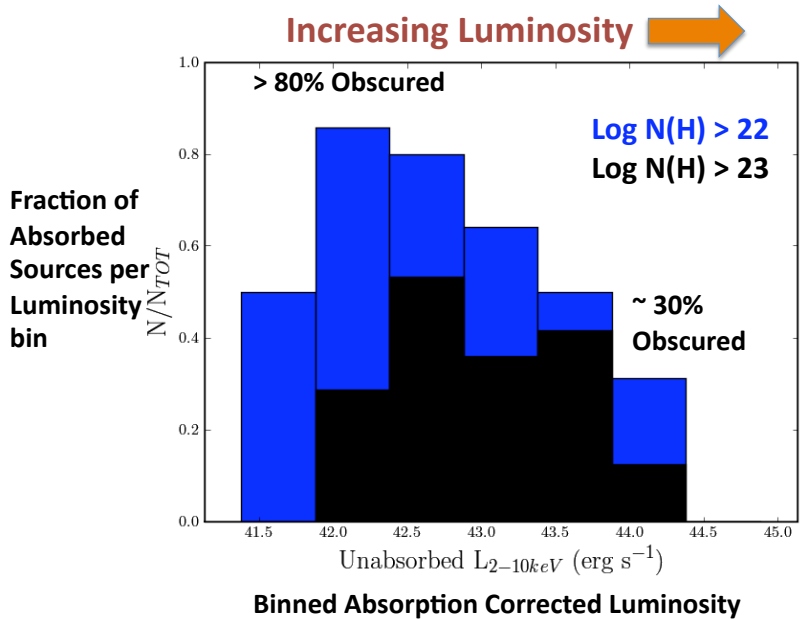}
  \caption{The X-ray analysis of the Swift-detected AGN shows that the fraction of absorbed sources (N$_{\rm H} > 10^{22}$\,cm$^{-2}$) increases at lower luminosities.  Conversely, there is a smaller fraction of absorbed sources at high luminosities.  Since the unified model predicts that unabsorbed and absorbed AGN are the same sources viewed at different angles, these results provide a challenge to the unified model. (Note that the lowest luminosity bin is not statistically significant due to few sources in this bin.)}
\end{figure}

As we mentioned, the most absorbed sources in the Swift sample have the most complex spectra.  They are also less luminous than the unabsorbed AGN.  In the following sub-sections, we discuss the properties of the most absorbed sources detected with Swift.  Many of these, the hidden AGN, were detected as AGN for the first time with Swift's BAT.  The elusive Compton-thick sources, the most obscured AGN found in nature, are also discussed.

\subsection{Hidden AGN}

Hidden or buried AGN were first identified through Suzaku follow-ups of Swift detected sources \citep{2007ApJ...664L..79U}.  Suzaku follow-ups were necessary, as little direct X-ray emission from these heavily absorbed sources is detected in the 0.3--10\,keV band.  Simultaneous spectra with the Suzaku XIS (sensitive to the soft through hard X-rays) and pin (sensitive $> 15$\,keV) allow for determinations of the basic X-ray properties, including the column density of obscuring gas.  The initial analysis of two hidden sources in \citet{2007ApJ...664L..79U} found that these are heavily obscured AGN with N$_{\rm H}$ from $5 \times 10^{23} - 10^{24}$\,cm$^{-2}$ that are likely buried in a geometrically thick torus.  Optical spectroscopy shows weak H$\beta$ and [O III] emission lines.

From our characterization of the X-ray properties of the entire Swift selected sample, we found that this class of AGN is significant.  We find that 24\% of local AGN are ``hidden''.  Analysis of the available 0.3--10\,keV spectra are ambiguous on whether these heavily obscured sources are Compton-thick.  Therefore, we obtained Suzaku spectroscopy of a sample of 5 hidden AGN to determine their properties.

The Suzaku spectra of the hidden sources, which were all detected as AGN for the first time with Swift, reveal that they are not Compton-thick \citep{2009ApJ...701.1644W}.  The combination of the Suzaku XIS and pin spectra reveal column densities from N$_{\rm H} = 4 - 10 \times 10^{23}$\,cm$^{-2}$.  The Compton thick signature of strong Fe K$\alpha$ emission lines, with $EW \sim 1$\,keV, are not observed.
Although they do have flat power-law indices ($\Gamma < 1.5$), another indicator of a reflection dominated spectrum.
Compton-thick sources are likely to show little variability since they are so heavily obscured that only reflected emission is viewed in the X-ray band.  We find that the hidden sources are very variable (see Figure 3), but that this variability is observed more in the hard/very hard X-rays than in the soft X-rays (where little emission is detected).  This is in contrast to unabsorbed sources, which are most variable in the soft X-rays.

Therefore, we find that hidden AGN are heavily obscured sources that are not Compton-thick.  This population is important, as we find that 24\% of the Swift-detected AGN are in this class.  However, the multi-wavelength properties are not well-known, as these sources are likely missed in optical/softer X-ray surveys.

\begin{figure}
  \includegraphics[height=.35\textheight]{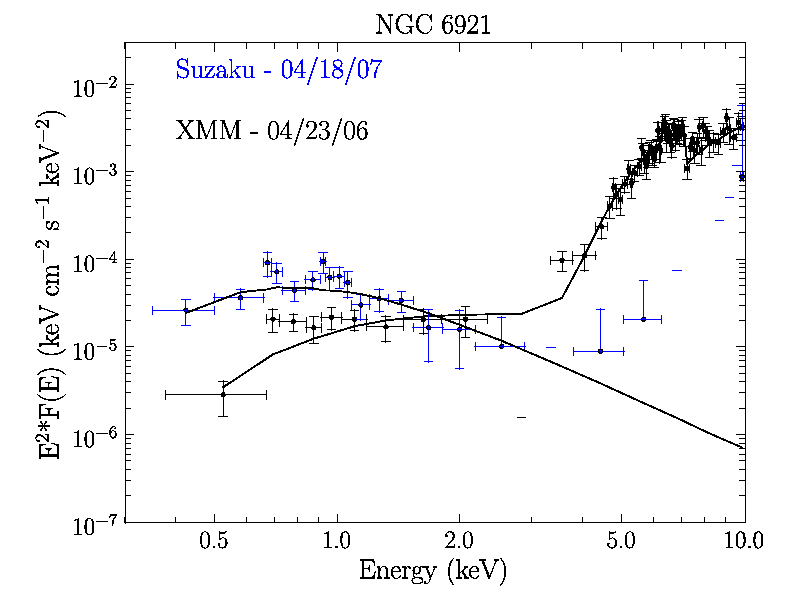}
  \caption{A detailed analysis of XMM-Newton and Suzaku spectra of hidden AGN \citep{2009ApJ...701.1644W} reveals that they are heavily obscured but Compton-thin.  The column densities are close to, but below the Compton-thick threshold.  Additionally, the spectra do not show strong Fe K$\alpha$ emission, a hallmark of reflected emission.  Further, the hidden sources are extremely variable, as the variability of NGC 6921, above, demonstrates.  Therefore, the hidden AGN, which make up 24\% of the Swift sample, are a population of heavily obscured, Compton-thin AGN.  The AGN nature of this new class is ``hidden'' in optical/softer X-ray surveys.
  }
\end{figure}

\subsection{Compton-thick AGN}

The detection of heavily obscured AGN is a main goal of the Swift AGN survey.  We find that the most heavily obscured AGN, Compton-thick sources, are difficult to detect even at the highest X-ray energies.  X-ray analysis of the 9-month survey reveal that a few Compton thick sources are detected, including well-known sources like Cen A, NGC 1275, and NGC 6240.  Characterizing the properties of such sources is difficult, however, as Compton-thick AGN are complicated.  Using discriminators of flat power-law indices, strong Fe K$\alpha$ $EW$ and high column densities, we find that 6\% of the Swift BAT-detected AGN are Compton-thick \citep{2009ApJ...690.1322W}.

Independent analysis of the Swift BAT spectra confirms this rate ($4.6^{+2.1}_{-1.5}$\%) \citep{2011ApJ...728...58B}.  However, \citet{2011ApJ...728...58B} determine that if the distribution of AGN is corrected for absorption, the corrected detection rate of Compton-thick AGN from Swift is much higher ($20^{+9}_{-6}$\%).  The Compton-thick rate is an important prediction for models of the cosmic X-ray background, which account for the observed 30\,keV emission with a significant fraction (10--15\%) of Compton-thick AGN   
\citep{2007AA...463...79G}.  While Swift is detecting some of these sources, we are still missing most of these AGN.  We must look to future very hard X-ray surveys, like NUSTAR, to uncover the most heavily obscured population.  However, in the softer X-rays, $< 10$\,keV, we find that the X-ray properties in the Swift sample perfectly account for the $\Gamma = 1.4$ slope of the CXB measured by 
\citet{1980ApJ...235....4M}.


\begin{figure}
  \includegraphics[trim = 1in 0.8in 1.5in 2.0in, clip, height=.4\textheight]{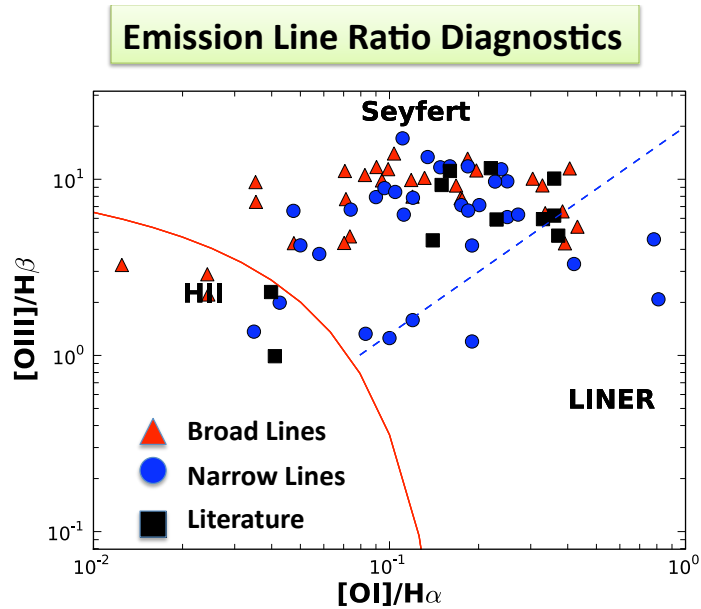}
  \caption{Optical emission line diagnostics, like the one shown, using the narrow component of prominent emission lines, reveal that the majority of the Swift-detected AGN are optical Seyferts.  We also detect a population of H II galaxies/composites/LINERs.  These sources are among the least luminous in the Swift sample.  When we compare the optically classified Seyfert 2s to the Seyfert 1s, we find that these sources have the same distribution of luminosities  Therefore, the unified model prediction of no difference in the luminosity of different AGN classes holds up for the optical Seyferts.}
\end{figure}

\begin{figure}
  \includegraphics[height=.3\textheight]{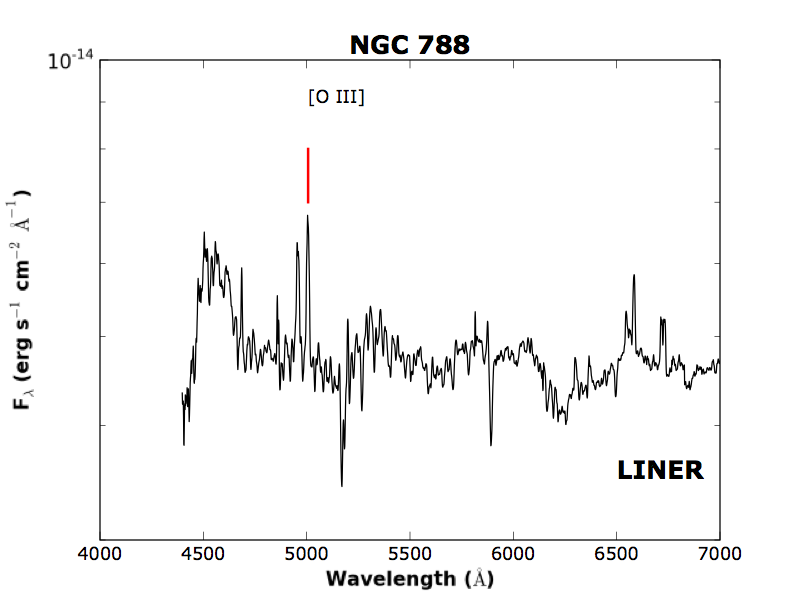}
  \hspace{-0.7cm}
  \includegraphics[height=.3\textheight]{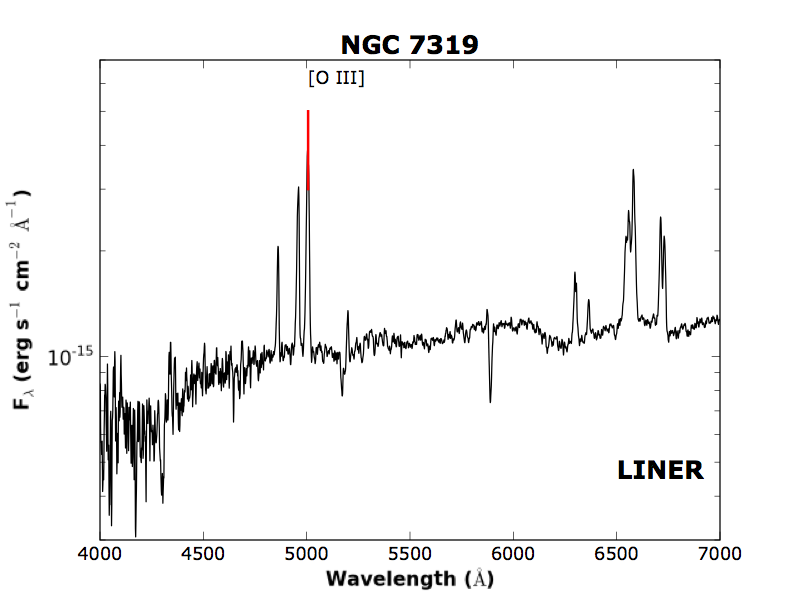}
  \caption{Many of the lowest luminosity/most absorbed AGN are not classified as optical Seyferts.  Shown are two examples of the optical spectroscopy of absorbed/low luminosity AGN that are LINERs.  The observed spectra are shown shifted to the AGN rest-frame.  These non-Seyfert sources are not consistent with the unified model.
  }
\end{figure}

\section{Clues to the Nature of Low Luminosity AGN from Optical Spectroscopy}\label{opt}

To determine the optical spectroscopic properties of the Swift very hard X-ray detected AGN, we analyzed spectra of our sources from the  Sloan Digital Sky Survey and our own follow-ups from the Kitt Peak National Observatory's 2.1-m telescope \citep{2010ApJ...710..503W}.  Our sample consisted of 81\% of the ``northern'' Swift-detected AGN.  To determine the optical emission line properties, we first modeled the continuum emission from the host galaxy using the stellar population models of \citet{2003MNRAS.344.1000B} and the method of 
\citet{2004ApJ...613..898T}.  The narrow and broad lines were decomposed, using Gaussian model fits to the continuum subtracted spectra.

We used the narrow emission-line diagnostics of \citet{1981PASP...93....5B} and \citet{1987ApJS...63..295V} (see Figure 4 for an emission line diagnostic plot).  We find that the narrow line sources consist of 66\% Seyferts, 16\% LINERs, 13\% ambiguous, 3\% composite H II/AGN, and 3\% H II galaxies.  The non-Seyfert sources and the LINERs in particular (examples shown in Figure 5), are more heavily absorbed.  The average X-ray column density for the LINERs is $6 \times 10^{23}$\,cm$^{-2}$, whereas the average column density for the absorbed Swift sources is $10^{23}$\,cm$^{-2}$.  Additionally, we find that the non-Seyferts are among the lowest luminosity sources in the sample.  If we look at the distribution of luminosities for the broad line AGN versus the narrow line Seyferts, we find that the luminosity distributions are consistent with being drawn from the same population.  Therefore, the luminosities of optical Seyferts are consistent with the unified model.  We find that the optical LINERs/composites/H II galaxies are not consistent.  These sources do not fit in the standard unified model picture and may be accreting in a different mode (e.g., radiatively inefficient accretion mode as discussed in \citet{2008ARA&A..46..475H}).  Many of these sources are classified in the X-rays as ``hidden'' AGN.

\section{Summary}

The Swift all-sky, very hard X-ray survey is uncovering a large sample of absorbed AGN that are missed in optical/softer X-ray surveys.  Among the new absorbed sources, we identify that 24\% of the very hard X-ray AGN are ``hidden''/buried.  Suzaku follow-ups, which allow for simultaneous coverage in the soft through very hard X-rays, of hidden AGN proved crucial for determining the column densities and basic X-ray properties of this class.  Hidden AGN are heavily obscured, Compton-thin sources.  Meanwhile, Swift finds that Compton-thick sources account for only $\sim 6$\% of the sample -- showing that these sources are faint even in the 14--195\,keV band.  

In the X-rays, absorbed AGN are less luminous than unabsorbed AGN.  However, optical spectroscopic follow-ups show that the absorbed sources classified as Seyferts have the same luminosity distributions as the optical broad line AGN.  The optical LINERs/H II galaxies/composites are among the most obscured and lowest luminosity sources.  These are the absorbed AGN that are not consistent with the unified model and are possibly accreting in a different mode than the Seyferts.  These results will continue to be tested through Suzaku and optical spectroscopic follow-ups of the additional 500+ AGN detected in the newest Swift catalog -- offering a new view on obscured AGN in the local Universe.


\begin{theacknowledgments}
  The author thanks the organizing committee of the {\it Exploring the X-ray Universe: Suzaku and Beyond}, Suzaku Science Conference, for inviting her to talk about the Swift AGN survey results.  Also, the author gratefully acknowledges the work of the Swift Gamma-ray burst satellite team (PI: Neil Gehrels), and the BAT survey team, in particular Jack Tueller and Richard Mushotzky.
\end{theacknowledgments}



\bibliographystyle{aipproc}   

\bibliography{/Users/lisa/Documents/Documents/MyBibtex.bib}


\end{document}